\documentclass[twocolumn,showpacs,preprintnumbers,amsmath,amssymb]{revtex4}
\usepackage{graphicx}
\begin{document}

\title{\bf
Anisotropy of incommensurate spin and charge fluctuations in
detwinned YBa$_2$Cu$_3$O$_{6+\delta}$ }

\author
{
 T. Zhou$^{1}$ and Jian-Xin Li$^{1,2}$
}
\address{$^{1}$National Laboratory of Solid State of Microstructure and
Department of Physics, \\ Nanjing University, Nanjing 210093, China\\
$^{2}$The Interdisciplinary Center of Theoretical Studies, Chinese
Academy of Science, Beijing 100080, China. }
\date{May 7, 2005}

\begin{abstract}
Motivated by a recent neutron scattering experiment on detwinned
YBa$_2$Cu$_3$O$_{6+\delta}$ superconductor, we examine the
frequency and doping dependence of the anisotropy in the spin and
charge fluctuation arising from the coupling between the plane and
the chain. Starting from the two-dimensional $t$-$t^{'}$-$J$ model
and using the random-phase approximation (RPA), we find a
pronounced anisotropy of the incommensurate (IC) peaks in the spin
channel, namely the peak intensity at the $(\pi\pm\delta,\pi)$
direction is stronger than that at the $(\pi,\pi\pm\delta)$
direction in a wide frequency range from $\omega=0.2J$ to the
resonance frequency $\omega_r=0.35J$. Above the resonance
frequency, the IC peaks reemerge. Their intensities shift to the
diagonal direction and no anisotropy exists along the two diagonal
directions. We find that this anisotropy is robust with respect to
the possible variation of the RPA correction factor and to the
dopings. The charge fluctuation is also found to be incommensurate
for all energies considered and peak at $(0,\delta)$ and
$(\delta,0)$. An anisotropy in its IC peak intensity along the
$k_x$ and the $k_y$ direction exists, but in sharp contrast to the
spin channel, the maximum intensity of the IC peak is along the
$k_y$ direction. Moreover, the IC peak in the charge channel
exhibits an upward dispersion, in contrast to the downward
dispersion below the spin resonance frequency for the spin IC
peak. We explain these features based on the effect of the
plane-chain coupling on the topology of the Fermi surface.

\end{abstract}


\maketitle

\section{Introduction}
The spin and charge dynamics are important issues in the studies
of the high-$T_c$ superconductivity. The inelastic neutron
scattering (INS) experiments have been playing important roles on
studying their spin dynamics. It can provide direct information of
the momentum and frequency dependence of the dynamical spin
susceptibility. On the other hand, the INS experiments can measure
the charge fluctuations indirectly by observing a change in the
mass density. In addition to the indirect measurement, the charge
susceptibility can be detected directly by the inelastic x-ray or
electron scattering.

Over the past decade, the INS experiments carried out on the
twinned samples have observed the four-fold symmetric
incommensurate (IC) peaks at the wave vector
${\bf{q}}=(\pi,\pi\pm\delta)$ and ${\bf{q}}=(\pi\pm\delta,\pi)$ in
a range of low energies~\cite{mas,moo,ros}, in addition to a sharp
commensurate resonance peak around the antiferromagnetic (AF) wave
vector ${\bf Q}=(\pi,\pi)$ at about 41 meV~\cite{tra,mo,fon}. The
experiment also reported that the IC peaks at $(0,\pm2\delta)$ are
observed for the charge excitation through the measurement of
atomic mass density change~\cite{ham,mdo}. Theoretically, there
are two possible explanations of the spin incommensuration in the
INS experiments. One is based on the striped-phase picture, which
suggests that the presence of the dynamic stripes is the origin of
the IC peaks~\cite{tran,eme}. The other is the scenario of the
nested-Fermi-surface which can explain the gross feature of the
incommensurate spin fluctuations and spin
resonance~\cite{bri,morr,kao,norman,jxli,li,jxl}.

Recently, Mook $\it et$ $al.$ have carried out the neutron
scattering measurement on partly detwinned
YBa$_2$Cu$_3$O$_{6.6}$~\cite{mook} and found that the IC peak
shows the one dimensional (1D) feature. They claimed that it gives
a strong support on the scenario of the stripe-phase picture.
However, we showed that~\cite{zho} an anisotropic IC peak can in
fact be obtained when the coupling between the plane and chain,
and the asymmetry of the crystal lattice between the $a$ and $b$
axis are considered in the two-dimensional $t$-$t^{'}$-$J$ model,
based on the Fermi surface topology. Very recently, an INS
experiment on fully detwinned YBCO~\cite{hin} shows that the
geometry of the magnetic fluctuations is actually two-dimensional,
but with a prominent anisotropy in the amplitude and the width of
the IC peak which is maximum along the $(q,\pi)$ direction and the
anisotropy is strongly energy-dependent. Our previous theoretical
calculation~\cite{zho} shows many features which are consistent
with the experimental data~\cite{hin}. However, the linewidth
anisotropy found in the experiment is more stronger than our
calculation at the fixed energy $\omega=0.3J$($J$ is the AF
coupling constant in the $t$-$t^{'}$-$J$ model) and a detail
theoretical investigation of the energy dependence of the
linewidth anisotropy as found experimentally still
lacks~\cite{note}. Another new development in recent INS
experiments is the reemergence of the IC peak at higher energies
above the spin resonance energy~\cite{hay,jmt}. This high-energy
IC peak is found to shift from the $(\pi,q)$(and its symmetric
points) direction to the diagonal direction~\cite{hay,jmt}. This
reemergence of the spin incommensuration has in fact been
predicted theoretically based on the scenario of the Fermi surface
topology~\cite{kao,norman,jxli}. However, both the experimental
and theoretical investigations are carried out in the twinned
systems. So, an interesting issue is how the IC peak behaves in
this high-energy region in the detwinned systems.

Though we have shown before that the plane-chain coupling, and the
asymmetry of the crystal structure can lead to the in-plane
anisotropy of the spin IC peak, their effect may differ in
detail~\cite{zho}. Namely, the former leads the maximum IC peak to
be along the $k_x$ direction, but the latter be along the $k_y$
direction. Due to the much stronger anisotropy caused by the
plane-chain coupling, the combined effect gives rise to a maximum
IC peak along the $k_x$ direction, which is consistent with
experiments~\cite{mook,hin}. Our investigation is based on a
self-consistent calculation of the quasiparticle dispersion and
the superconducting (SC) gap in the framework of the slave-boson
approach to the $t$-$t^{'}$-$J$ model. We note that a recent
calculation based on the similar RPA-type theory with the
phenomenological quasiparticle dispersion and SC gap obtained by
fitting to experiments shows that the asymmetric crystal structure
also leads to the $k_x$-direction IC peak~\cite{ere}, in contrast
to what we obtained before. We find that this contrasting behavior
comes from the different relative ratios between the $a-b$
anisotropy of the nearest-neighbor hopping constant and the SC gap
asymmetry, both are caused by the orthorhombic crystal structure.
The asymmetric SC gap leads to a $k_y$-direction IC peak, while
the anisotropy in the hopping constant leads to a $k_x$-direction
IC peak. Our self-consistent calculation gives a
$\Delta_x/\Delta_y\approx1.5$ (the gap function is $\Delta_{\bf
k}=\Delta_x \cos k_x-\Delta_y \cos k_y$), which is bigger than the
value used by Eremin {\it et al.}~\cite{ere}. So, the overwhelming
asymmetric SC gap in our calculation leads to the anisotropy be in
the $k_y$ direction. Therefore, the explanation of the anisotropy
in the IC peak based on the orthorhombic crystal structure may
depend on the fine tuning of the dispersion and SC gap, in
particular its effect is shown to be much weaker than that from
the plane-chain coupling~\cite{zho}. So, we will just focus on the
effect of the plane-chain coupling in this paper.

In this paper, we will first extend our previous scheme~\cite{zho}
to address the above issues in view of the new experimental
results~\cite{hin,hay,jmt}. We will also examine in detail the
effects on the anisotropic IC peaks of dopings and the RPA
correction factor $\alpha$(see the text below)~\cite{bri}, in
order to check the robustness of this anisotropy. Our aim is not
only to account for the new findings observed by the very recent
experiment~\cite{hin}, but also to predict some new features which
help to determine if the scenario of the nested-Fermi-surface can
describe all features of the spin incommensuration or some exotic
scenarios such as the stripe phase or the electronic nematic
phase~\cite{kao1} may be required. In addition, we will also
calculate the charge susceptibility and investigate the frequency
dependence of the anisotropy in the charge channel.


The article is organized as follows. In Sec. II, we introduce the
model and define notations. The Hamiltonian and formalism are
obtained. In Sec. III, we present numerical results of the spin
and charge susceptibility, respectively. In Sec. IV, we interpret
the results. Finally, we present the conclusion in Sec. V.

\section{Hamiltonian and Formalism}

We start with a Hamiltonian which describes the system with a
CuO$_2$ plane and a CuO chain per unit cell~\cite{zho}.
\begin{equation}
H=H_{t-t^{'}-J}+H_c+H_I,
\end{equation}
where $H_{t-t^{'}-J}$ is the two-dimensional $t$-$t^{'}$-$J$
Hamiltonian describing the CuO$_2$ plane,
\begin{eqnarray}
H_{t-t^{'}-J}&=&-t\sum_{i\alpha\sigma}
c^{\dagger}_{i\sigma}c_{i+\alpha\sigma}-h.c.-t^{'}\sum_{i\sigma}c^{\dagger}_{i\sigma}c_{i+\hat{x}+\hat{y}\sigma}-h.c
\nonumber\\&&+J\sum_{i\alpha}({\bf{S}}_i\cdot{\bf{S}}_{i+\alpha}-\frac{1}{4}n_in_{i+\alpha}),
\end{eqnarray}
$H_c$ describes the CuO chain,
\begin{equation}
H_c=-\sum_{i\alpha\sigma}t_cd^{\dagger}_{i\sigma}d_{i+\alpha\sigma}-h.c.,
\end{equation}
and the coupling between the plane and chain is~\cite{xia},
\begin{eqnarray}\label{hi}
H_I&=&-t_{\perp}\sum_{ij\sigma}(c^{\dagger}_{i\sigma}d_{j\sigma}+h.c)
-{\lambda}/4\sum_{i\alpha}(\widehat{\Delta}^{\dagger}_{1,i,\alpha}\widehat{\Delta}_
{2,i+\bar{c}/2,\alpha}\nonumber\\&&+\widehat{\Delta}^{\dagger}_{1,i,\alpha}\widehat{\Delta}_{2,i-\bar{c}/2,\alpha}+h.c),
\end{eqnarray}
where $c^{\dagger}_{i\sigma}$ and $d^{\dagger}_{i\sigma}$ are the
creation operators of electrons in the plane and chain,
respectively, $\alpha=\hat{x},\hat{y}$ stands for unit lattice
vector along the $a$ and $b$ axis, and
${\bf{S}}_i=\frac{1}{2}c^{\dagger}_{i\alpha}\sigma_{\alpha\beta}c_{i\beta}$.
$\widehat{\Delta}_{nr{\alpha}}$ are the singlet pair operators,
which are expressed by
$\widehat{\Delta}_{1r{\alpha}}=c_{r\uparrow}c_{r+\alpha\downarrow}-c_{r\downarrow}c_{r+\alpha\uparrow}$
and
$\widehat{\Delta}_{2r{\alpha}}=d_{r\uparrow}d_{r+\alpha\downarrow}-d_{r\downarrow}d_{r+\alpha\uparrow}$.
Following Ref.~\cite{zho}, we use the slave-boson mean-field
theory to the plane Hamiltonian $H_{t-t^{'}-J}$, in which the
physical electron operators $c_{i\sigma}$ are expressed by slave
bosons $b_i$ carrying the charge and fermions $f_{i\sigma}$
representing the spin, $c_{i\sigma}=b^{\dagger}_if_{i\sigma}$. The
mean-field Hamiltonian can be written in a matrix
form~\cite{atk,zho}, and the mean-field order parameters
$\Delta_{1ij}=<f_{i\uparrow}f_{j\downarrow}-f_{i\downarrow}f_{j\uparrow}>=\pm\Delta_1$,
$\Delta_{2ij}=<d_{i\uparrow}d_{j\downarrow}-d_{i\downarrow}d_{j\uparrow}>=\pm\Delta_2$
($\pm$ depend on if the bond $<ij>$ is in the $\hat{x}$ or the
$\hat{y}$ direction),
$\chi_{ij}=\Sigma_{\sigma}<f^{\dagger}_{i\sigma}f_{j\sigma}>=\chi_0$,
and the chemical potentials $\mu_1$ and $\mu_2$ for different
planar doping densities $x$ can be obtained from the
self-consistent equations~\cite{zho}. The doping densities for the
chain are fixed to $x_{ch}=0.5$. The other parameters are choosed
as $t=2J$, $t^{'}=-0.45t$, $\lambda=3J$, the renormalized
plane-chain hoping constant
$\widetilde{t}_\perp=t_\perp\sqrt{x}=-0.1J$, and the temperature
$T=0.005J$. The chain is along $k_y$ direction.

The bare spin susceptibility $\chi^{s}_0({\bf q},\omega)$ coming
from the particle-hole excitations of the planar fermions can be
calculated from the two-particle Green's function $\chi^{s}_0({\bf
q},i\omega)$ by the analytical continuation
$i\omega\rightarrow\omega+i\delta$,
\begin{eqnarray}\label{chi0}
\chi^{s}_0({\bf
q},\omega)&=&\frac{1}{N}\sum_{\bf{k}}\sum_i\sum_j[U^{2}_{1i}({\bf{k}})U^{2}_{1j}({\bf{k}}+{\bf{q}})
\nonumber\\&&+U_{1i}({\bf{k}})
U_{2i}({\bf{k}})U_{1j}({\bf{k}}+{\bf{q}})U_{2j}({\bf{k}}+{\bf{q}})]\nonumber\\&&
\times\frac{f(E_j({\bf{k}}+{\bf{q}}))-f(E_i({\bf{k}}))}{\omega-E_j({\bf{k}}+{\bf{q}})+E_i({\bf{k}})+i\delta}.
\end{eqnarray}
with $U_{ij}$ are the elements of the matrix which diagonalizes
the mean-field Hamiltonian~\cite{zho}.

The correction of the AF spin fluctuations to the spin
susceptibility is included in the form of the renormalized
random-phase approximation (RPA),

\begin{equation}
\chi^{s}({\bf q},\omega)=\chi^{s}_0({\bf
q},\omega)/[1+{\alpha}J_{\bf q}\chi^{s}_0({\bf q},\omega)].
\end{equation}
Here $J_{\bf q}=J(\cos{q_x}+\cos{q_y})$. The RPA correction factor
$\alpha$ ($\alpha<1$) is introduced to overcome the overestimation
of the AF correlation in the usual ($\alpha=1$)
RPA~\cite{bri,yma}. Experimentally, the spin resonance peak is
observed at 41meV at the optimal doping $x=0.14$ which corresponds
to about $0.35J$. Thus, whenever not be indicated, the AF
correction factor $\alpha=0.5$ will be chosen to make the
resonance frequency occur at $0.35J$ at the optimal doping density
$x=0.14$.

The bare charge susceptibility can also be calculated from the
two-particle Green's function,
\begin{eqnarray}\label{chi1}
\chi_0^{c}({\bf
q},\omega)&=&\frac{1}{N}\sum_{\bf{k}}\sum_i\sum_j[U^{2}_{1i}({\bf{k}})U^{2}_{1j}({\bf{k}}+{\bf{q}})
\nonumber\\&&-U_{1i}({\bf{k}})
U_{2i}({\bf{k}})U_{1j}({\bf{k}}+{\bf{q}})U_{2j}({\bf{k}}+{\bf{q}})]\nonumber\\&&\times
\frac{f(E_j({\bf{k}}+{\bf{q}}))-f(E_i({\bf{k}}))}{\omega-E_j({\bf{k}}+{\bf{q}})+E_i({\bf{k}})+i\delta}.
\end{eqnarray}
In the form of RPA, the renormalized charge susceptibility can be
written as
\begin{equation}
\chi^{c}({\bf q},\omega)=\chi^{c}_0({\bf
q},\omega)/[1-\frac{1}{4}J_{\bf q}\chi^{c}_0({\bf q},\omega)].
\end{equation}

\section{the numerical results}
\subsection{dynamical spin susceptibility}

\begin{figure}

\centering

\includegraphics[scale=0.5]{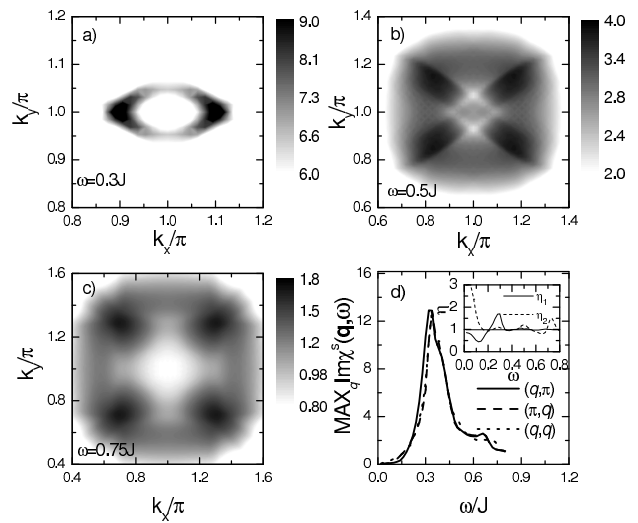}

\caption{Panels a)-c) are the intensity plot of the imaginary part
of the spin susceptibility Im$\chi^{s}$ as a function of the wave
vector with doping $x=0.14$ for $\omega=0.3J$, $0.5J$, and
$0.75J$, respectively. Panel d) shows the frequency dependence of
the maximal intensity of Im$\chi^{s}$. The inset of panel d) shows
the ratio of the maximal Im$\chi^{s}$ at different momentum
directions (see text).} \label{fig1}
\end{figure}

Fig.1(a-c) show the intensity plot of the imaginary part of the
spin susceptibility in the momentum space with doping $x=0.14$ at
frequencies $0.3J, 0.5J,$ and $0.75J$, respectively. The IC peak
intensity as a function of frequency at different momentum
directions is shown in Fig.1(d). From Fig.1(a), one can see that
the IC peaks show clear anisotropy between the $k_x$ and $k_y$
direction at $\omega=0.3J$, namely, the incommensurability
$\delta$ and the intensity of the IC peaks at
$(\pi\pm\delta_x,\pi)$ is obviously larger than that at
$(\pi,\pi\pm\delta_y)$. When frequency is increased to be above
the resonance frequency $0.35J$, the IC peaks rotate to the
diagonal direction, as is shown in Fig.1(b) and Fig.1(c).
Interestingly, now no anisotropy exists along the two diagonal
directions. The anisotropy below the resonance frequency is found
to be frequency dependent, as shown in Fig.1(d). Detailed
frequency dependence can be found in the inset of Fig.1(d), in
which the ratio of the peak amplitude between the $k_x$ and $k_y$
direction ($\eta_1$), and between the diagonal and the $k_y$
direction ($\eta_2$), as a function of energy $\omega$ is
presented
$(\eta_1(\omega)=$max$_q$Im$\chi^{s}(q,\pi,\omega)/$max$_q$Im$\chi^{s}(\pi,q,\omega),
\eta_2(\omega)=$\\
max$_q$Im$\chi^{s}(q,q,\omega)/$max$_q$Im$\chi^{s}(\pi,q,\omega))$.
As shown, the anisotropy is significant in a wide frequency range
from slightly below the resonance frequency to about $0.2J$. It
increases firstly when the energy is reduced from the resonance
and reaches the maximal value $\eta_1$=1.7 when frequency is at
about 0.28J, then decreases with the further decrease of energy.
The frequency dependence of the anisotropy is reasonably
consistent with the experiment~\cite{hin}, in which the maximal
ratio is about 2.1 at $\omega=28$ meV. Below $0.2J$, the IC peak
rotates again to the diagonal direction as expected from the
node-to-node excitations at low frequencies.

\begin{figure}

\centering

\includegraphics[scale=0.5]{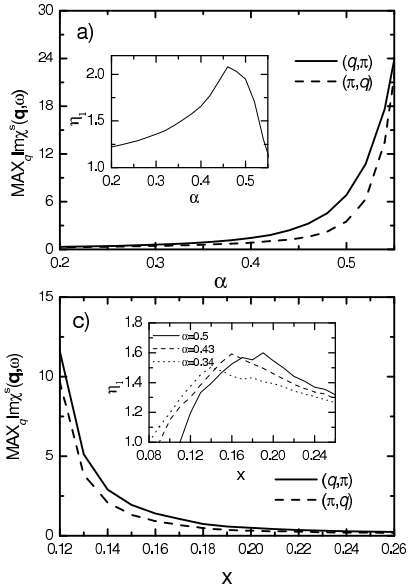}

\caption{Dependence of the maximal peak in Im$\chi^{s}$ on the RPA
correction factor $\alpha$ [a)] and doping $x$ [b)], with
$\omega=0.3J$. The inset of figures shows the ratio of the maximal
Im$\chi^{s}$ between the $(q,\pi)$ and the $(\pi,q)$ direction.}
\label{fig2}
\end{figure}

We now study the dependence of the anisotropy on the possible
variations of the AF correction factor $\alpha$ with frequency
$\omega=0.3J$ and on doping density $x$ with frequency
$\omega=0.25J$ in Fig.2(a-b). The main panel shows the maximal
intensity of spin excitations at different momentum directions and
the inset is the ratio of the peak amplitude between the $q_y=\pi$
and the $q_x=\pi$ direction($\eta_1$). From Fig.2(a), one can see
that the anisotropy is enhanced gradually with an increasing of
the AF correction factor $\alpha$. When $\alpha$ equals 0.47, the
anisotropy reaches its maximal value. Then, the further increase
of $\alpha$ makes the anisotropy decrease and when $\alpha=0.57$,
the anisotropy vanishes. As shown in Eq.(6), the AF correction
factor $\alpha$ acts to enhance the AF correction, so the
resonance peak is shifted to lower energies with the increase of
$\alpha$. When $\alpha=0.57$, the spin resonance right appears at
$\omega=0.3J$, so the anisotropy vanishes because the spin
resonance peaks at $(\pi,\pi)$. The critical doping density for
the AF instability observed by experiments is $x_c=0.02\sim0.05$.
It will give rise to $\alpha=0.34\sim0.43$ if one determine the AF
correction factor $\alpha$ according to this experimental value
$x_c$~\cite{bri,jxli,li}. In this range of $\alpha$, the
anisotropy is about $1.35\sim 2.0$, as can be seen from the inset
of Fig.2(a), so it is robust with respect to the possible
variation of $\alpha$. The doping dependence of the anisotropy
with frequency $\omega=0.25J$ is shown in Fig.2(b). Remarkably,
the anisotropy increases with doping in the underdoped regime,
then decreases with doping and still have a ratio $\eta_{1}=1.2$
at the heavily overdoped regime $x=0.26$. This specific behavior
is due to the variation of the topology of the Fermi surface with
doping. Because this dependence is experimentally accessible, it
will be interesting to compare the experimental result with our
theoretical prediction, considering that the recent experiment was
just carried out at the nearly optimal doped
YBa$_{2}$Cu$_{3}$O$_{6.85}$~\cite{hin}.

\begin{figure}

\centering

\includegraphics[scale=0.5]{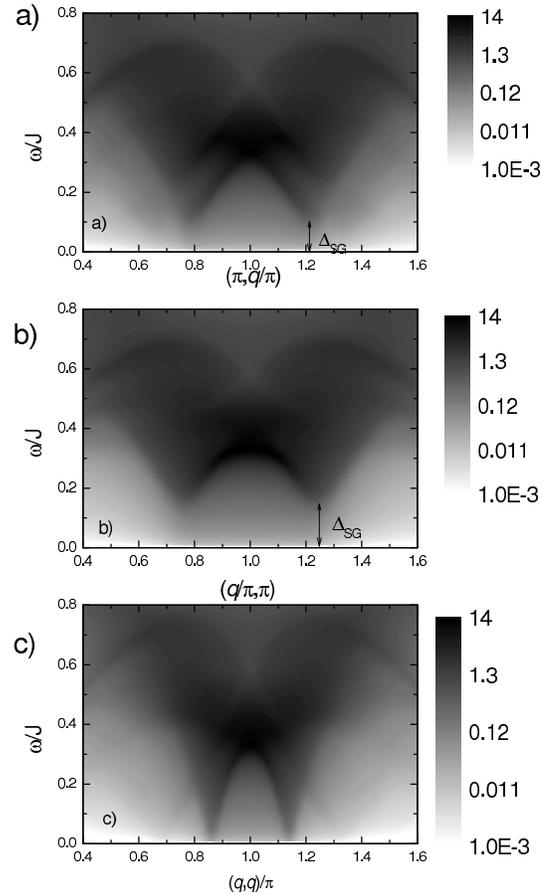}

\caption{Intensity plot of Im$\chi^{s}$ as a function of frequency
$\omega$ and wave vector $\bf q$. Panels a)-c) are along the
$(\pi,q)$, $(q,\pi)$ and $(q,q)$ direction, respectively.}
\label{fig3}
\end{figure}

Fig.3(a-c) show the intensity plot of the imaginary part of the
spin susceptibility Im$\chi^{s}$ as a function of the frequency
and momentum. At the resonance frequency, the peak position is at
$(\pi,\pi)$. Below and above the resonance frequency the IC peaks
develop. The incommensurability is bigger as the frequency is
gradually far away from the resonance frequency. Therefore, they
exhibits a downward dispersion for frequency below the resonance
frequency and an upward dispersion for frequency above the
resonance frequency. This feature is consistent with the recent
neutron scattering data~\cite{hay,jmt}. From Fig.3(a) and
Fig.3(b), we can see an obvious spin gap at the $(q,\pi)$ and
$(\pi,q)$ direction. Compared with that calculated for the system
without the plane-chain coupling~\cite{jxli}, a pronounced
anisotropy in the spin gap is found, namely, the spin gap in the
$(q,\pi)$ direction is bigger than that in the $(\pi,q)$
direction. Due to the existence of nodes, there is no spin gap
along the diagonal direction as shown in Fig.3(c).

\subsection{dynamical charge susceptibility}

\begin{figure}

\centering

\includegraphics[scale=0.5]{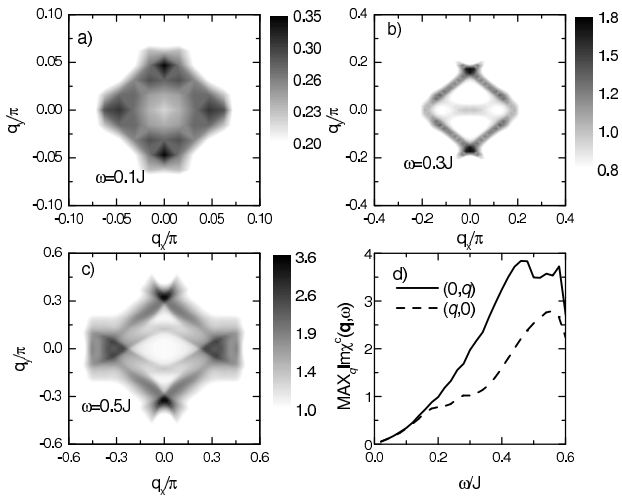}

\caption{Panels a)- c) are the intensity plot of the imaginary
part of the charge susceptibility Im$\chi^{c}$ as a function of
wave vector with doping $x=0.14$ for $\omega=0.1J$, $0.3J$ and
$0.5J$, respectively. Panel d) shows the frequency dependence of
the maximal intensity of Im$\chi^{c}$. The solid line denotes the
result along the $(0,q)$ direction and the dashed line along the
$(q,0)$ direction.} \label{fig4}
\end{figure}

In Fig.4(a-c), we show the imaginary part of the charge
susceptibility Im$\chi^{c}$ as a function of the momentum with
doping $x=0.14$ at frequencies 0.1J, 0.3J, and 0.5J, respectively.
An obvious feature of the charge susceptibility is that its IC
peak appears at small momenta around $(0,0)$, namely, at
$(0,\delta)$ and $(\delta,0)$. However, in sharp contrast to the
anisotropy in the spin channel, the maximum IC peak in the charge
channel is along the $k_y$ direction. Meanwhile, the IC peak is
always along the $(0,\delta)$ and $(\delta,0)$, while that in the
spin channel will rotate to the diagonal direction at low and high
frequencies. The frequency dependence of the IC peak intensity is
shown in Fig.4(d). As shown, the anisotropy exists in a wide
frequency range. It firstly appears at $\omega_c=0.1J$ and becomes
significant above $\omega=0.2J$.

\begin{figure}

\centering

\includegraphics[scale=0.5]{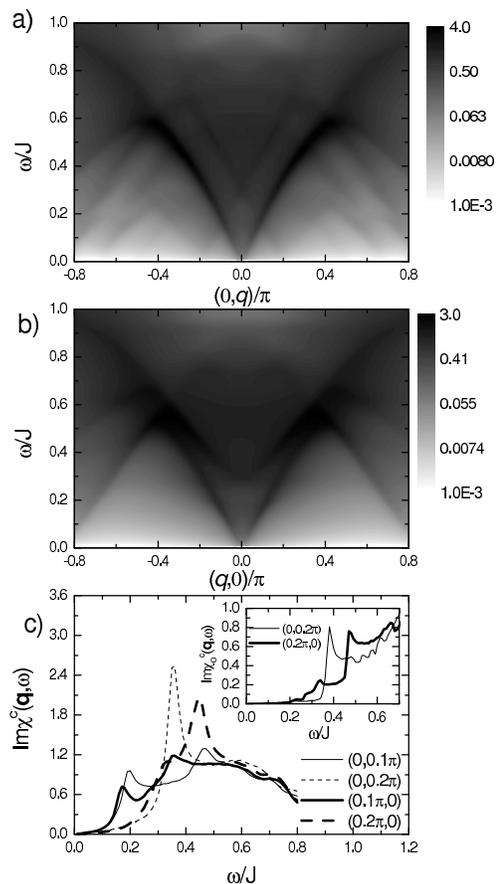}

\caption{Panels a) and b) are the intensity plot of Im$\chi^{c}$
as a function of frequency $\omega$ and wave vector $\bf q$ along
$(0,q)$ and $(q,0)$ direction, respectively. Panel c) is the
frequency dependence of Im$\chi_c$ with different wave vectors.
The inset of panel c) shows the frequency dependence of
Im$\chi^{c}_0$.} \label{fig5}
\end{figure}

The intensity of the charge susceptibility as a function of the
momentum and frequency is shown in Fig.5(a) and Fig.5(b). The
charge susceptibility is incommensurate for all frequencies
considered. The incommensurability will increase as the frequency
increases and form a usual upward dispersion. This is quite
different from the dispersion found in the spin channel, where the
spin resonance at the commensurate AF wave vector ${\bf
Q}=(\pi,\pi)$ disparts the dispersion into a downward form at low
frequencies and an upward form at high frequencies. In the
stripe-phase picture for the spin IC fluctuation~\cite{tran}, a
change in the AF phase of the stripe at the charge-domain boundary
occurs and it leads to a spin periodicity twice that of the charge
periodicity. As a result, a two-to-one ratio of the
incommensurability $\delta$ between the charge fluctuation and the
spin fluctuation is expected. Here, though we can obtain the
relation at a certain frequency $\omega=0.3J$, generally there
should be no such correspondence because of the different
curvature of their dispersion. This feature comes from the
topology of the Fermi surface and its origin will be discussed in
the following section. It constitutes an additional difference
between the Fermi-surface explanation and the stripe-phase
explanation. In Fig.5(c), the imaginary part of the charge
susceptibility as a function of the frequency is shown at small
momentum transfers. For smaller momentum transfer ${\bf q}$, the
charge excitation exhibits a broad spectrum. With the increase of
${\bf q}$, the lower energy spectrum is suppressed and a peak
occurs at the edge of the gap. With the increase of ${\bf q}$, the
peak intensity increases and reaches the maximum intensity around
$|{\bf q}|=0.4\pi$, as can be seen from Fig.5(a) and (b). From the
inset of Fig.5(c) we can see that the imaginary part of the bare
charge susceptibility also has a peak at the same frequency.
Moreover, we have checked the pole equation $1-1/4J_{\bf
q}\chi^{c}_0({\bf q},\omega)=0$, and found that it does not
satisfy at the peak energy. So, this peak is not the resonance
peak as observed in the spin channel, rather it is right the IC
charge peak. Thus, an anisotropy in the peak frequency and width
exists in two different directions of the momentum transfer ${\bf
q}$, as shown in Fig.5(c).

\section{discussion}

\begin{figure}

\centering

\includegraphics[scale=0.5]{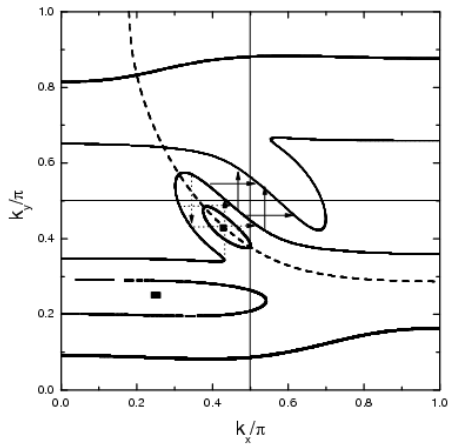}

\caption{Contour plot of the quasiparticle energy $E_{\bf
k}=\omega/2$ shown in the first quadrant of the Brillouin zone for
$\omega=0,0.12J,0.3J$ and its by $(\pi,\pi)$ shifted image at
$\omega=0.3J$. The dashed line shows the normal state Fermi
surface. The solid arrows denote the nesting vectors relative to
$(\pi,\pi)$. The dotted arrows denote the nesting vectors within
the energy contour. } \label{fig6}
\end{figure}

Let us now explain the origin of the above features based on the
topology of the Fermi surface. In the Fermi-liquid based theory,
the spin and charge susceptibilities arise from the scattering of
the SC quasiparticle across the SC gap. Their only difference is
the sign difference in their coherence factors, namely,
$c^{s(c)}=1-(\varepsilon_{\bf k}\varepsilon_{{\bf k}+{\bf
q}}\pm\Delta_{\bf k}\Delta_{{\bf k}+{\bf q}})/E_{\bf k}E_{{\bf
k}+{\bf q}}$, which is reflected by the plus/minus sign in front
of $U_{1i}$ in Eq.(5) and (7). Therefore, the coherence factor is
appreciable when the gap $\Delta_{\bf k}$ and $\Delta_{{\bf
k}+{\bf q}}$ have the opposite sign for the spin channel and the
same sign for the charge channel. Thus, the peak of the imaginary
part of the spin susceptibility will be near the AF wave vector
${\bf Q}=(\pi,\pi)$ because $\Delta_{\bf k}=-\Delta_{{\bf k}+{\bf
Q}}$ and that of the charge susceptibility will be near the
$(0,0)$ point. Then, the origin of the observed IC peaks can be
illustrated in term of the initial and final scattering processes
along energy contours, with different transfer wave vectors for
the spin and the charge channel. Fig.6 shows the contour plot of
the quasiparticle energy in the SC state $E_{\bf k}=\omega/2$ at
$\omega=0,0.12J,0.3J$, and its by $(\pi,\pi)$ shifted image at
$\omega=0.3J$. In the spin channel, the transfer wave vector of
the main scattering processes should be near $(\pi,\pi)$.
Therefore, for an intermediate excitation energy such as
$\omega=0.3J$, the best nesting vector (solid arrows in Fig.6)
will be a horizontal and vertical offset to $(\pi,\pi)$, which
connects the flat part of the energy contour with its by
$(\pi,\pi)$ shifted image. The scattering with this best nesting
vector contributes directly to the IC peaks, therefore the IC
peaks will appear in the parallel direction. In the case that
there is no plane-chain coupling, the energy contour will be
symmetric with respect to the diagonal direction. The coupling
between the plane and chain will break this symmetry and the
resulting nesting portion of the Fermi surface connected by the
nesting vector along the $k_x$ direction is larger than that along
the $k_y$ direction, as shown in Fig.6. So, both the imaginary and
the real part of the bare spin susceptibility $\chi^{s}_0$ show an
obvious anisotropy, i.e., the peak intensity at
$(\pi\pm\delta_x,\pi)$ is stronger than that at
$(\pi,\pi\pm\delta_y)$.

\begin{figure}

\centering

\includegraphics[scale=0.5]{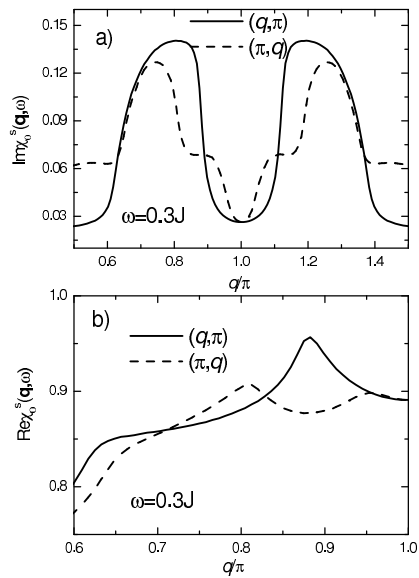}

\caption{Panel a) shows the imaginary part of the bare spin
susceptibility as a function of wave vector with $\omega=0.3J$.
Panel b) is its real part.} \label{fig7}
\end{figure}

However, the anisotropy coming just from Im$\chi^{s}_0$ is in fact
small, as can be seen from a comparison between that in Fig.7(a)
and Fig.1(d). The RPA correction factor $1+\alpha
J_{\bf{q}}$Re$\chi^{s}_0$ in the renormalized spin susceptibility
Im$\chi^{s}$ acts to enhance this anisotropy. The minimum of the
RPA correction factor along the $(q,\pi)$ direction is less than
that along the $(\pi,q)$ direction, because the maximum value of
Re$\chi^{s}_0$ along the former direction is larger than the
latter direction, as shown in Fig7(b). As frequency approaches to
the resonance frequency, Re$\chi^{s}_0$ will tend to $-1/(\alpha
J_{\bf{q}})$ and this makes the IC peak move close to $(\pi,\pi)$.
Therefore, the anisotropy will become less and less and eventually
vanish at the resonance energy. In addition, the RPA correction
factor also affects strongly the incommensurability $\delta$ of
the spin susceptibility. From Fig.7 (a), one can notice that the
IC peak position in Im$\chi^{s}_0$ along the
$(\pi\pm\delta_{x_0},\pi)$ direction is nearer to the $(\pi,\pi)$
point than that along the $(\pi,\pi\pm\delta_{y_0})$ direction
($\delta_{x_0}<\delta_{y_0})$. While in the renormalized spin
susceptibility [shown in Fig.1(a)], this anisotropy of the
incommensurability is reversed, i.e., $\delta_x>\delta_y$. In the
experiment~\cite{hin} one can also find that the peak position at
$(\pi,\pi\pm\delta_y)$ is a little nearer to $(\pi,\pi)$ than that
at $(\pi\pm\delta_x,\pi)$.

At low frequencies, the RPA correction factor plays minor role.
The anisotropy of the spin gap along $k_x$ and $k_y$ direction is
mainly determined by the anisotropy of the bare spin
susceptibility. As frequency tends to zero, only node-to-node
excitation becomes available, so there is no spin excitation along
the $(q,\pi)$ and $(\pi,q)$ direction. As the frequency increases
to $\omega_y=0.12J$, the constant energy contour crosses the line
$k_x=\pi$, as shown in Fig.6. So, the spin excitation along
$(\pi,q)$ appears because now this constant energy contour and its
by $(\pi,\pi)$ shift image [not shown in Fig.6] can be connected
by the nesting wave vector along the $k_{y}$ direction. The
critical frequency $\omega_y$ may be taken approximately as the
spin gap along the $k_y$ direction. Because the constant energy
contour does not cross the line $k_y=\pi$ at this frequency, there
is still no spin excitation along the $k_x$ direction, i.e., the
spin gap along the $k_x$ direction is bigger than that along the
$k_y$ direction.

In the high energy region, the energy contour will evolve into two
constant energy lines which are parallel approximately to the
Fermi surface in the normal state, so the best nesting wave vector
will rotate to the diagonal direction. In this case, the nesting
wave vector along the $(\pi+\delta, \pi+\delta)$ and the
$(\pi+\delta, \pi-\delta)$ direction is symmetric. Thus, the
plane-chain coupling can not cause the anisotropy in the diagonal
direction in the Fermi-liquid based theory.

In the charge channel, the transfer wave vector of the main
scattering processes should be near $(0,0)$. Therefore, the best
nesting vector (dotted arrows in Fig.6) is within the energy
contour and contributes the IC peak at $(0,\delta)$ and
$(\delta,0)$. At low energies, the plane-chain coupling is less
effective, so the constant energy contour is basically symmetric
with respect to the diagonal direction and there is no anisotropy
in the charge fluctuation. As the frequency increases, the
plane-chain coupling destructs part of the nesting portion. It
leads the nesting portion of the Fermi surface along the $k_y$
direction to be larger than the $k_x$ direction, so the IC peak
along the $k_y$ direction is stronger than that along the $k_x$
direction.

\section{conclusion}
In summary, we examine the frequency evolution of the spin and
charge susceptibility based on the two dimensional $t$-$t'$-$J$
model considering the coupling between the plane and chain. In the
spin channel, the IC peaks appear at $(\pi,\pi\pm\delta_y)$ and
$(\pi\pm\delta_x,\pi)$ below the resonance
frequency($0.2J<\omega<0.35J$). In this region, a pronounced
anisotropy of the spin susceptibility is found, i.e., the peak
intensity along the $k_x$ direction is stronger than that along
the $k_y$ direction. Above the resonance frequency, the IC peaks
reemerge and rotate to the diagonal direction, and there is no
anisotropy in the IC peak intensity along the two diagonal
directions. The dependence of the anisotropy on the
antiferromagnetic correction factor $\alpha$ and the doping
density $x$ is investigated, and it is found that the anisotropy
is robust in the reasonable range. In addition, we also find an
obvious anisotropy of the spin gap between the $(\pi,q)$ and the
$(q,\pi)$ direction. The spin gap along the $k_x$ direction is
bigger than that along the $k_y$ direction. In the charge channel,
the susceptibility is also incommensurate for all energies we
considered. The IC peaks appear at $(0,\pm\delta)$ and
$(\pm\delta,0)$. In sharp contrast to the spin channel, the IC
peak intensity in the charge channel along the $k_y$ direction is
stronger than that along the $k_x$ direction. Meanwhile, the IC
peak exhibits an upward dispersion in the whole energy region, in
contrast to the downward dispersion below the resonance frequency
in the spin channel. In addition, no resonance mode is found in
the charge channel. We explain all the results based on the
scenario of the nesting Fermi surface and suggest that the
coupling between the plane and chain is responsible for the
observed anisotropy.

\begin{acknowledgments}
This work was supported by National Nature Science Foundation of
China (10474032,10021001) and by RFDP (20030284008).

\end{acknowledgments}

\end{document}